\documentclass[aps,pre,superscriptaddress,nofootinbib]{revtex4}
\pdfoutput=1
\usepackage[a4paper, margin=2.0cm]{geometry}
\usepackage{graphicx}
\usepackage{bm}

\usepackage{tablefootnote}
\usepackage{amsmath}
\usepackage{amssymb}
\usepackage{url}
\newcommand{\e}[1]{\ensuremath{\times 10^{#1}}}
\usepackage{lmodern}
\usepackage[T1]{fontenc}
\usepackage{placeins}

\usepackage[hidelinks]{hyperref}

\usepackage[hidelinks]{hyperref}

\hypersetup{
    pdfauthor={Nathan Clisby},
    pdftitle={Monte Carlo study of four-dimensional self-avoiding walks of up to 1 billion steps}
    }

\providecommand{\norm}[1]{\vert #1 \vert}
\providecommand{\ee}[1]{\ensuremath{\times 10^{#1}}}

\newcommand{\Z}{{\mathbb Z}}
\newcommand{\resq}{R_{\mathrm{E}}^2}

\newcommand{\rgsq}{R_{\mathrm{G}}^2}

\newcommand{\dde}{D_{\mathrm{E}}}
\newcommand{\ddg}{D_{\mathrm{G}}}
\newcommand{\Nmin}{N_{\text{min}}}
\newcommand{\avresq}{\langle R_{\mathrm{E}}^2 \rangle}
\newcommand{\avrgsq}{\langle R_{\mathrm{G}}^2 \rangle}
\newcommand\tstrut{\rule{0pt}{2.4ex}}
\newcommand\bstrut{\rule[-1.0ex]{0pt}{0pt}}
\begin{document}

\title{Monte Carlo study of four-dimensional self-avoiding walks of up
to one billion steps}

\author{Nathan Clisby} 

\affiliation{School of Mathematics and Statistics,
	The University of Melbourne, 
        Victoria 3010, Australia}

\email{nclisby@unimelb.edu.au}

\date{March 31, 2017}

\begin{abstract}
We study self-avoiding walks on the four-dimensional hypercubic lattice
via Monte Carlo simulations of walks with up to one billion steps. 
We study the expected logarithmic corrections to scaling, and find 
convincing evidence in support the scaling form predicted by the
renormalization group, with an estimate for the power of the logarithmic
factor of 0.2516(14), which is consistent with the predicted value of 1/4.
We also characterize the behaviour of the pivot algorithm for sampling 
four-dimensional self-avoiding walks, and conjecture that the probability of
a pivot move being successful for an $N$-step walk is $O([ \log N ]^{-1/4})$.
\end{abstract}

\maketitle

\section{Introduction}
\label{sec:intro}

Self-avoiding walks (SAWs) are the set of walks on a graph, typically
the integer lattice $\Z^d$, where each step is between nearest
neighbours, and no vertex is visited twice. The model has a long
history, and has played a pivotal role in our understanding of polymers
and critical phenomena more broadly, and in the development of conformal
field theory, the lace expansion, and Schramm-Loewner
Evolution~\cite{Madras1993SelfAvoidingWalk,Bauerschmidt2012LecturesSelfAvoidingWalks}.

In three dimensions SAWs are in the same universality class as real
polymers, and provide an ideal laboratory for studying universal
properties such as critical exponents.

In four dimensions the model is no longer physically relevant, but
nonetheless it is worthy of study as it provides a useful test of
renormalization group techniques which have made various predictions. It
can also be regarded as a test-bed problem for physically relevant
models which have logarithmic corrections to scaling, for example the
$\theta$-transition in three dimensions is believed to be identified
with a tricritical point, and has mean-field behavior with logarithmic
corrections. Four-dimensional SAWs have been studied by Monte
Carlo~\cite{Grassberger1994Selfavoidingwalks,Owczarek2001ScalingSelfAvoiding}
and
enumeration~\cite{MacDonald1992Selfavoidingwalks,Chen2003AmplitudeRatiosSelf,Clisby2007Selfavoidingwalk}
methods, and we note that rigorous results have recently been obtained
for the 4-dimensional weakly self-avoiding walk via the rigorous
renormalization group~\cite{Brydges2010Renormalisationgroupanalysis}.

In this work we will restrict ourselves to the study of SAWs on $\Z^4$.
The key quantities of interest are the number of SAWs of $N$ steps,
which we denote $c_N$, and the mean size of SAWs of length $N$.
We can formally describe a walk of $N$ steps as a mapping $\omega$ from the
integers $0,1,\cdots,N$ to sites on $\Z^4$, with $\norm{\omega(i+1) -
\omega(i)} = 1$ $\forall i\in[0,N-1]$, and $\omega(i) \neq \omega(j)$
$\forall i \neq j$.
We calculate the two most common measures of size, the squared
end-to-end distance, $\resq$, and the squared radius of gyration,
$\rgsq$, which are defined as:
\begin{align}
    \label{eq:defre2}
    \resq & = \norm{\omega(N) - \omega(0)}^2;
\\
\label{eq:defrg2}
    \rgsq & = \frac{1}{2 (N+1)^2} \sum_{i,j}
    \norm{\omega(i)-\omega(j)}^2.
\end{align}
Our goal is to understand the asymptotic behavior of the mean squared
end-to-end distance and mean squared radius of gyration, in the
long-chain limit.
Renormalization group
arguments~\cite{Duplantier1986PolymerChainsInFourDimensions}
give the following asymptotic forms~\cite{Duplantier1986PolymerChainsInFourDimensions,Grassberger1994Selfavoidingwalks} for the relevant quantities in four
dimensions:
\begin{align}
\label{eq:cn} 
c_N &= A \mu^N [\log (N/a)]^{1/4} \left(1 -
\frac{17\log(4\log(N/a))-3}{64\log(N/a)} + \cdots \right); \\
\label{eq:re2} 
\avresq_N &=  D_{\rm E} N [\log (N/a)]^{1/4} \left(1 -
\frac{17\log(4\log(N/a))+31}{64\log(N/a)} + \cdots \right); \\
\label{eq:rg2}
\avrgsq_N &=  D_{\rm G} N [\log (N/a)]^{1/4} \left(1 -
\frac{17\log(4\log(N/a))+97/3}{64\log(N/a)} + \cdots \right).
\end{align}
Here, $\langle \cdots \rangle_N$ indicates the expectation of the
observable over the set of SAWs of length $N$, $\mu =
6.774\;043(5)$\cite{Owczarek2001ScalingSelfAvoiding} is the growth
constant, $A$, $\dde$, and $\ddg$ are lattice dependent amplitudes, and
$a$ is a common scale factor.  The overall factor of $[\log N]^{1/4}$ 
in Eqs~\ref{eq:re2} and \ref{eq:rg2} is
universal, and would occur for any observable which measures the squared
size of the walk, such as the square of the hydrodynamic radius, and the
mean squared monomer-to-end distance.  While the amplitudes themselves
are not universal, it is expected that the amplitude ratio $\dde/\ddg$
is a universal quantity. For SAWs in four dimensions this should have
the same value as for simple random walks, i.e. we should have
$\dde/\ddg = 6$. 

The logarithmic corrections arise because $d=4$ is the upper critical
dimension for self-avoiding walks. For $d \geq 5$ self-avoiding walks
are in the same universality class as simple random
walks~\cite{Hara1992SelfAvoidingWalk,Hara1992LaceExpansionSelf}, or
Brownian motion, where there are no logarithmic corrections and, for
example, $\avresq_N = O(N)$.

Logarithmic corrections are notoriously hard to observe, but recent
developments in Monte Carlo simulation methods have made it conceivable
that we could reach the asymptotic regime and carefully test the
asymptotic forms given in Eqs~\ref{eq:re2} and \ref{eq:rg2}. The pivot
algorithm~\cite{Lal1969MonteCarlocomputer,Madras1988PivotAlgorithmHighly}
is a Markov chain Monte Carlo algorithm which is the most efficient
known method for sampling SAWs of fixed length. Recent
improvements~\cite{Kennedy2002fasterimplementationpivot,Clisby2010AccurateEstimateCritical,Clisby2010Efficientimplementationpivot}
have reduced the CPU time necessary to attempt a pivot move to $O(\log
N)$, which makes it possible to reach the regime of truly large $N$,
where the limit is due to computer memory (RAM) availability, rather
than any constraint of available computer time.

We have applied the SAW-tree
implementation~\cite{Clisby2010Efficientimplementationpivot} of the
pivot algorithm for the
first time to the problem of sampling four-dimensional self-avoiding walks, and so
generated high precision estimates of $\avresq_N$ and $\avrgsq_N$ for
SAWs of up to one billion steps. In the process we also studied the
behaviour of the SAW-tree implementation, and the characteristics of the
pivot algorithm for sampling self-avoiding walks.

In the remainder of this paper, we first describe our Monte Carlo
simulations in Sec.~\ref{sec:mc}, which includes data on the performance
of the SAW-tree implementation, then analyse the results in
Sec.~\ref{sec:analysis}, and conclude in Sec.~\ref{sec:conclusion}.

\section{Monte Carlo simulation}
\label{sec:mc}

The pivot algorithm is a Markov chain Monte Carlo algorithm that samples
SAWs of fixed length $N$. The basic move is a pivot, which involves the
application of a lattice symmetry (rotation or reflection) to one piece
of a walk around a chosen site of the walk. A pivot move is successful
if it results a walk that is self-avoiding, and so generates a correlated
sequence of self-avoiding walk configurations. In the seminal work of
Madras and Sokal~\cite{Madras1988PivotAlgorithmHighly}, the pivot
algorithm was proved to sample SAWs uniformly at random, and it was also
shown to be remarkably efficient at sampling global observables, such as
$R_{\rm E}^2$, due to the fact that with each successful pivot move a
large change is made to global observables. As mentioned in the
introduction, recent
improvements~\cite{Kennedy2002fasterimplementationpivot,Clisby2010AccurateEstimateCritical,Clisby2010Efficientimplementationpivot}
have increased the efficiency of the pivot algorithm still further, to
the point that it is now possible to sample SAWs with $10^9$ steps.
See~\cite{Madras1988PivotAlgorithmHighly} for many more details about
the pivot algorithm.

For the present computer experiment, we chose pivot sites uniformly at
random along the chain, and the pivot symmetry operations were chosen
uniformly at random from amongst the 383 lattice symmetries of $\Z^4$
that do not correspond to the identity.

To initialize the system we used the pseudo\_dimerize procedure, and to
eliminate any initialization bias we then performed approximately $20 N$
successful pivots.

For the longest walks, with $N=10^9$, this initialization procedure took
approximately 360 hours (or 15 days) to complete. To reduce this
computational burden, we performed an additional trick. We generated
seed walks for lengths from $10^6$-$10^9$; we expect these walks to be
indistinguishably close to equilibrium. We then used these seed walks
for separate simulations (8, in the case of $N=10^9$) where instead of
performing $20 N$ successful pivots to warm up the system we only
performed approximately $N$ successful pivots. Even though these are not
enough pivot moves to bring an initial configuration such as a straight
rod to equilibrium, given that we are at equilibrium this is more than 
sufficient to ensure
that correlations between estimates of global observables between
batches are completely negligible. 

After $20 N$ successful pivots had been applied for walks with $N < 10^6$,
or $N$ successful pivots had been
applied to seed walks for $N \geq 10^6$, we started collecting data for
our observables for each time step, and aggregated the results in
batches of $10^8$.

We sampled self-avoiding walks on $\Z^4$ over a range of lengths from
two thousand to one billion steps. We used the SAW-tree implementation
of the pivot algorithm as described
in~\cite{Clisby2010Efficientimplementationpivot} to collect data for the
observables $\avresq$, $\avrgsq$, the probability of a pivot move being
successful, $f$, and the amount of CPU time for each batch.
The computer experiment was run for 130 thousand CPU hours on Dell
PowerEdge FC630 machines with Intel Xeon E5-2680 CPUs.
In total there were $4.4 \times 10^5$
batches of $10^8$ attempted pivots, and thus there were a grand total of
$4.4 \times 10^{13}$ attempted pivots across all walk sizes.

Our data for $\avresq$,$\avrgsq$, $f$, and mean CPU time per pivot
attempt, are collected in Tables~\ref{tab:rdata} and \ref{tab:fdata}
in Appendix~\ref{sec:data}. We also include there the estimates for
the universal amplitude ratio $\avresq/\avrgsq$, as the positive
correlation between
two observables results in variance reduction
in the final estimates. That is, the error for the ratio is smaller than
would naively be expected from the original observables.

\begin{figure}[!htb]
  \begin{center}
    \includegraphics[width=9cm]{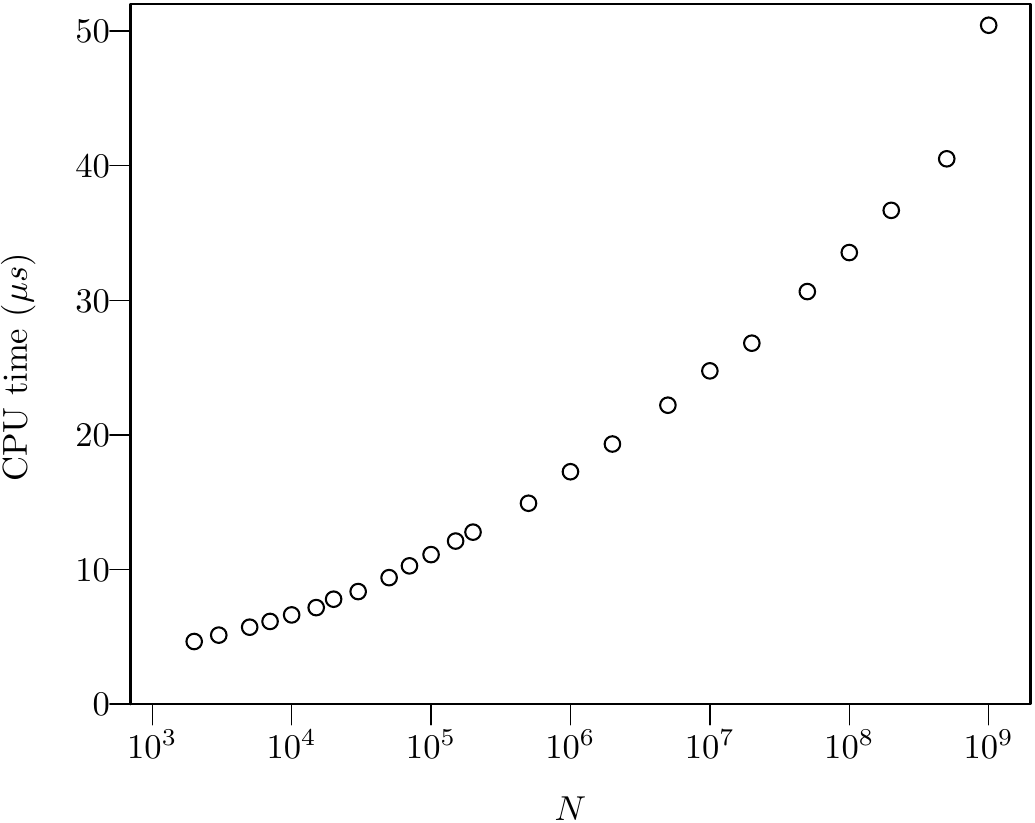}
  \end{center}
    \caption{CPU time per attempted pivot move as a function of $N$.}
  \label{fig:cpu}
\end{figure}

We plot the mean CPU time per attempted pivot against $N$ in
Fig.~\ref{fig:cpu}, with a logarithmic scale on the $x$-axis. 
The small amount of visible scatter should not be taken too seriously, as the computers
executing the code obey scheduling algorithms which mean that the CPU
time to run computations may vary somewhat. But, the overall trend is
consistent with the behavior demonstrated in our earlier
work~\cite{Clisby2010Efficientimplementationpivot} on SAWs in two and
three dimensions: the CPU time
increases logarithmically with $N$, although there is some degradation
in performance at large $N$, which is likely due to the larger memory
demand, which in turn means that memory accesses are more frequently out
of cache.

Interestingly, 
the equivalent plots in~\cite{Clisby2010Efficientimplementationpivot}
demonstrated quite starkly the impact of memory cache on performance by
exhibiting a clear kink, whereas here the degradation is far more
gradual. It is plausible that this effect is due to changes in CPU
architecture over the past seven years, i.e. since the earlier computer
experiment was performed.

In absolute terms, the performance is exceptionally fast: each pivot
attempt takes, on average, approximately 5 $\mu s$ for $N$ of the order
of thousands, and only takes ten times as long, or roughly $50 \mu s$, for walks of
one billion steps.

\section{Analysis}
\label{sec:analysis}

Our method of analysis is to perform
weighted non-linear fits of our data to the presumed asymptotic form, by
using the `nls' routine of the statistical programming
language R.
We truncated our data by only fitting values
with $N \geq N_\mathrm{min}$. We then varied $N_\mathrm{min}$ to get a sequence
of estimates for which we expect the systematic error due to unfitted
corrections-to-scaling to decrease.
We varied $\Nmin$ from 2000, which is the smallest value of $N$ for
which data were collected, to 
$\Nmin = 2\e{6}$, which is the largest value for which we felt
that the parameter estimates were informative (i.e. with sufficiently
small error bars that they aided in extrapolation).

We plotted our estimates against a variable which is of the same
relative size as the first neglected correction-to-scaling term.
In principle, this should result in linear convergence in the
plots as the asymptotic regime is reached, although it may be too
optimistic to hope for linear convergence in our case, as the neglected
correction terms are $O([\log N]^{-1})$ and $O([\log N]^{-2})$, which
suggests that some curvature would likely still be apparent even for
$N=10^9$. We extrapolate the fits from the right to where they intersect
the $y$-axis, which corresponds to the $N_{\mathrm{min}} \rightarrow
\infty$ limit.  There is a significant degree of subjectivity in
performing these extrapolations, so we present all of our fits
graphically to allow readers to judge for themselves whether the
extrapolations are suitably cautious.

We make the general observation that such analysis is fraught with
danger, especially when one ``knows'' the answer. It is exceedingly easy
to exhibit confirmation bias~\cite{Nickerson1998ConfirmationBiasReview}
when selecting appropriate analysis methods. Given that various
methods of analysis are available, and also variants of the same
method\footnote{E.g., in our case we may choose which how many terms to
include in our fits, whereas for differential approximant analyses of series it is possible to
vary the order of the differential equation.}, it is natural that some of
them will result in estimates which are closer to the expected answer.
From there it is a slippery slope to the conclusion that somehow the
methods which give the ``correct'' answer are the right methods
for the problem at hand. At this point it is possible that only the
results from the preferred method are presented in a research article,
so introducing bias, or, equally insidiously, it may be that the
confidence intervals for the preferred methods are estimated to be
smaller than for the non-preferred ``wrong'' methods.

This is exactly our situation, as the asymptotic forms for $\avresq$ and
$\avrgsq$ given in Eqs~\ref{eq:re2}
and \ref{eq:rg2} respectively are widely believed to be correct.
We have endeavored to reduce the likelihood of confirmation bias occurring by
including estimates from all of the methods of analysis which have been
used. We are fortunate that the power of the leading logarithm, found
in Eqs~\ref{eq:re2} and \ref{eq:rg2}, occurs for two observables, which
naturally gives us two somewhat independent estimates.

We now proceed with our analysis, starting with the raw data for
$\avresq$ and $\avrgsq$ from Table~\ref{tab:rdata}, to which we will fit the asymptotic forms in 
Eqs~\ref{eq:re2} and \ref{eq:rg2}. We start by fitting the leading
behavior only, where we assume that there is indeed a factor of $N$, but
where we fit for the amplitude $D$, parameter $a$, and the power of the
logarithm, which we denote as $\kappa$. That is, 
our statistical models are:
\begin{align}
    \avresq &= \dde N [\log (N/a)]^{\kappa}; \\
    \avrgsq &= \ddg N [\log (N/a)]^{\kappa}.
\end{align}

We find that this fitting form reproduces the Monte Carlo data quite
well for both $\avresq$ and $\avrgsq$, giving reduced $\chi^2$ values of
approximately 15-20 for $\Nmin =
2000$, and declining to approximately 1 for to $\Nmin \geq 5\e{5}$.
The fact that the statistical models are capable of reproducing the data
to within our error bars is evidence in support of the correctness of
the expressions given in 
Eqs~\ref{eq:re2} and \ref{eq:rg2}.

\begin{figure}[htb]
\begin{center}
\begin{minipage}{0.45\textwidth}
\begin{center}
    \includegraphics[width=1.0\textwidth]{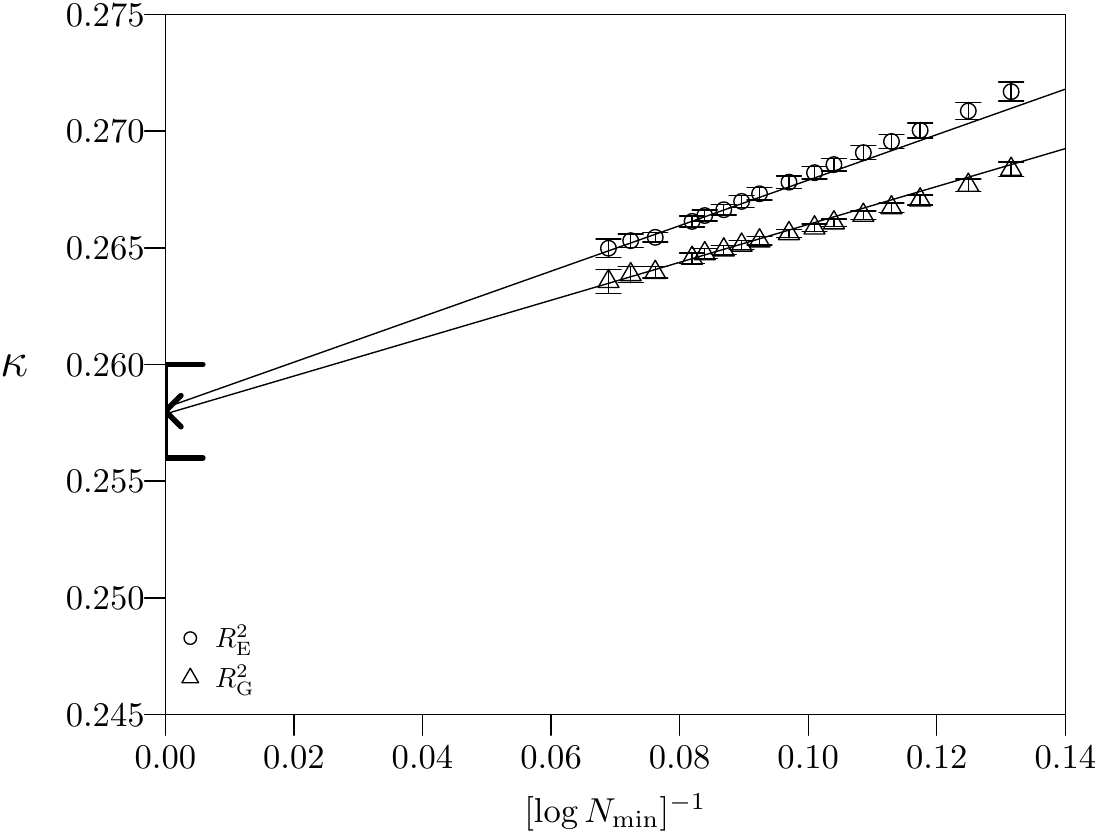}
\end{center}
\vspace{-4ex}
  \caption{Variation of fitted value of $\kappa$ with
    $N_{\rm min}$ when only the leading term is fitted.
    The lines of best fit to the final six values are shown,
    and our final estimate is plotted on the $y$-axis.}
  \label{fig:kappa}
\end{minipage}
\hspace{2em}
\begin{minipage}{0.45\textwidth}
\begin{center}
    \includegraphics[width=1.0\textwidth]{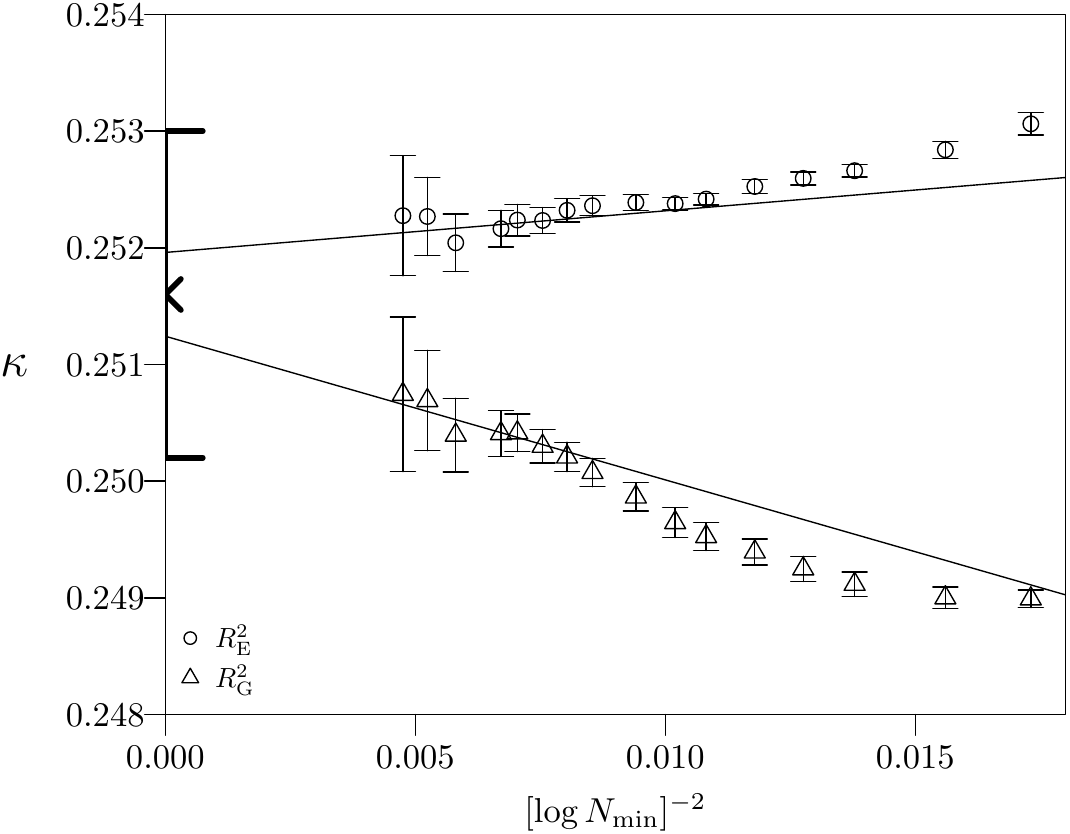}
\end{center}
\vspace{-4ex}
  \caption{Variation of fitted value of $\kappa$ with
    $N_{\rm min}$ when correction term is fitted.
    The lines of best fit to the final six values are shown,
    and our final estimate is plotted on the $y$-axis.}
  \label{fig:kappacorrection}
\end{minipage}
\end{center}
\end{figure}

We show the estimates of $\kappa$ from our fits in Fig.~\ref{fig:kappa}.
The estimates appear to smoothly converge to a value that is above
$1/4$, and our best estimate from extrapolating the trend is $\kappa =
0.258(2)$, which is shown on the $y$-axis of the plot. If taken at face
value this which would seem to exclude $\kappa = 1/4$, but we know that
the unfitted corrections to scaling may be large and so are mindful that
this may be misleading. (Admittedly, if we did not have prior
information that $\kappa = 1/4$, or of the details of the
correction-to-scaling terms, then the smoothness of the trend would have
given no hints that anything was amiss, and we would have been happy in
that case to report $\kappa = 0.258(2)$ as our final estimate.)

We note that we find that estimates for $a$ vary dramatically between
fits. In principle, it should have the same value regardless of
observable, and regardless of the number of terms used in the fits, but
in practice it behaves as a free parameter and shows no sign of
convergence to a consistent value. This was already observed by
Grassberger et al.~\cite{Grassberger1994Selfavoidingwalks}.  
We will not report in full the values obtained, but note that for
the fits of $\avresq$ the estimates of $a$ change monotonically
from 2.56 to 3.53 for the leading order fits as $\Nmin$ is increased,
from 0.201 to 0.207 for the fits with a correction term, and from 0.219
to 0.231 for the fits with a correction term and biasing $\kappa=1/4$.
The corresponding trends for the same fits for $\avrgsq$ are 3.54 to
4.45, 0.273 to 0.254, and 0.265 to 0.263 respectively (the last sequence
of estimates is not monotonic). 

We now fit the correction term, where we once again assume that
there is indeed a factor of $N$, but fit $D$, parameter $a$, and
$\kappa$. Our statistical models are:
\begin{align}
    \avresq &= 
D_{\rm E} N [\log (N/a)]^{\kappa} \left(1 -
    \frac{17\log(4\log(N/a))+31}{64\log(N/a)} \right); \\
    \avrgsq &= 
D_{\rm G} N [\log (N/a)]^{\kappa} \left(1 -
    \frac{17\log(4\log(N/a))+97/3}{64\log(N/a)} \right).
\end{align}

We find that this fitting form reproduces the Monte Carlo data extremely
well for both $\avresq$ and $\avrgsq$.  For $\avresq$, the value of 
reduced $\chi^2$ is only 4.5 for $\Nmin = 2000$, and hovers around
1 for $\Nmin \geq 15000$. For $\avrgsq$, the reduced $\chi^2$ is only
3.7 for $\Nmin = 2000$, and hovers around 1.3 for $\Nmin \geq 70000$.
We feel that the most likely explanation for the  spectacular fit for
$\avresq$ is happenstance, due perhaps to fortuitous cancellation
between competing neglected correction-to-scaling terms over the range
of $N$ that is fitted.

We show the estimates of $\kappa$ from our fits in
Fig.~\ref{fig:kappacorrection}. There we observe that the estimates are
much closer to $1/4$, and if we extrapolate the trends we obtain an
estimate of $\kappa = 0.2516(14)$ which is plotted on the $y$-axis. The
value of 1/4 is only just outside our confidence interval, and we
conclude that $\kappa = 1/4$ is consistent with these fits.

We now proceed to obtain
estimates of the amplitudes by biasing our fits based on the assumption that
$\kappa = 1/4$.
Once again we include the correction term, 
and in this case our statistical models are:
\begin{align}
    \avresq &= 
D_{\rm E} N [\log (N/a)]^{1/4} \left(1 -
    \frac{17\log(4\log(N/a))+31}{64\log(N/a)} \right); \\
    \avrgsq &= 
D_{\rm G} N [\log (N/a)]^{1/4} \left(1 -
    \frac{17\log(4\log(N/a))+97/3}{64\log(N/a)} \right).
\end{align}

We find that this fitting form reproduces the Monte Carlo data quite
well for both $\avresq$ and $\avrgsq$, but not as well as when $\kappa$
was allowed to vary.  For $\avresq$, the value of the
reduced $\chi^2$ is 31 for $\Nmin = 2000$, and declines to 2.3 for
$\Nmin = 2\e{6}$. For $\avrgsq$, the reduced $\chi^2$ is 
11 for $\Nmin = 2000$, and hovers around 1.5 for $\Nmin \geq 50000$.
Now, we see from Fig.~\ref{fig:kappacorrection}
that for $\avrgsq$ that the unbiased estimates for $\kappa$ are very
close to $1/4$ over a wide range of values for $\Nmin$, and so it makes
sense that the effect of biasing $\kappa = 1/4$ degrades the fits for
$\avrgsq$ to a lesser extent than for $\avresq$.

\begin{figure}[htb]
\begin{center}
\begin{minipage}{0.45\textwidth}
\begin{center}
    \includegraphics[width=1.0\textwidth]{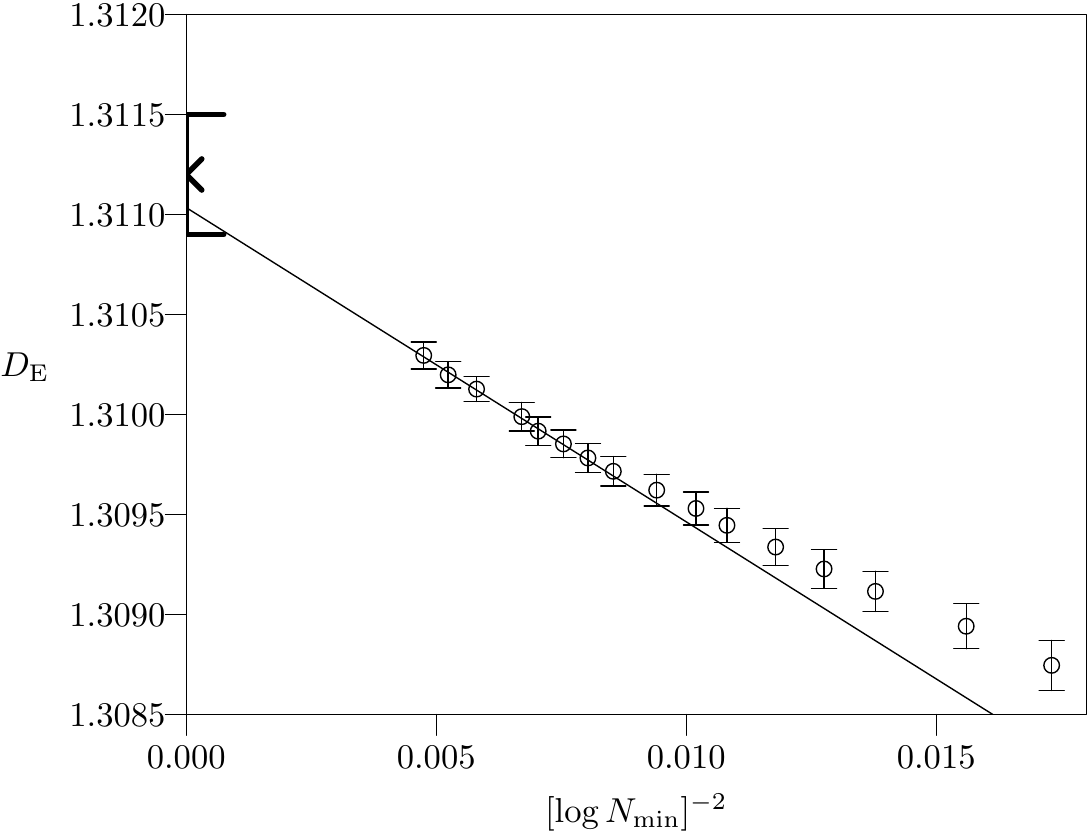}
\end{center}
\vspace{-4ex}
  \caption{Variation of fitted value of $\dde$ with
    $N_{\rm min}$ when correction term is fitted.
    The line of best fit to the final six values is shown,
    and our final estimate is plotted on the $y$-axis.}
  \label{fig:DE}
\end{minipage}
\hspace{2em}
\begin{minipage}{0.45\textwidth}
\begin{center}
    \includegraphics[width=1.0\textwidth]{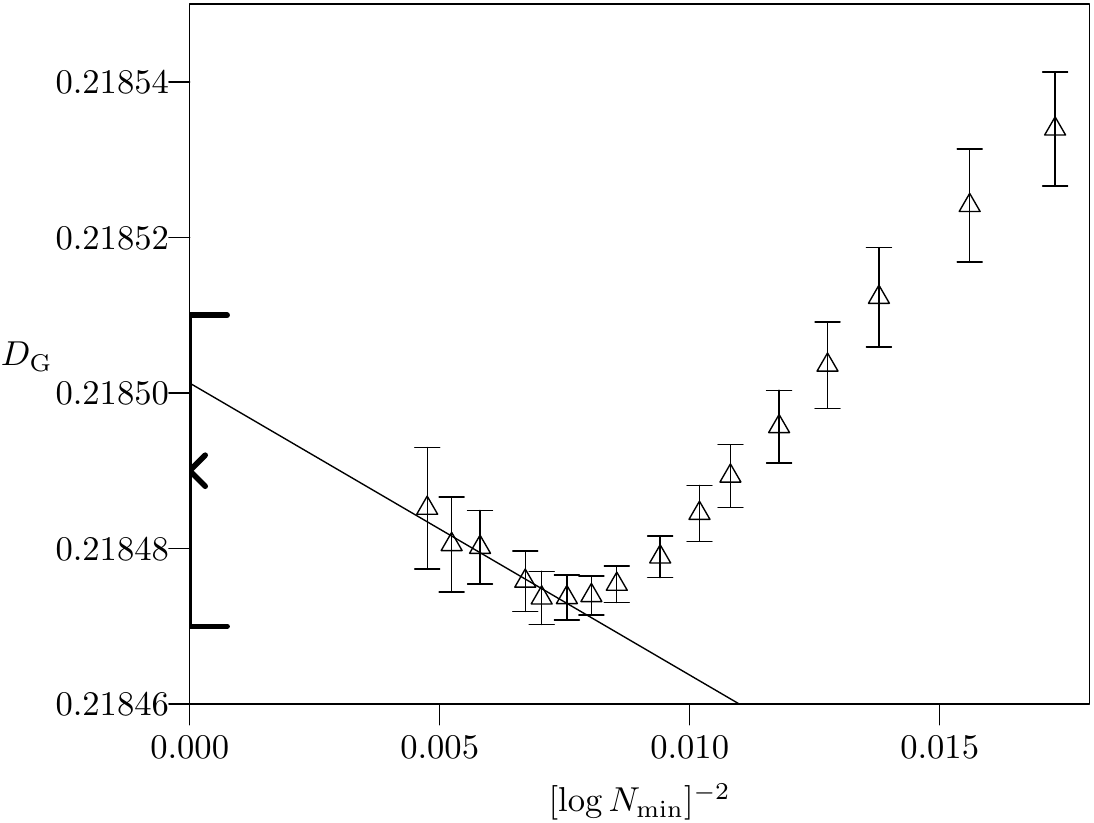}
\end{center}
\vspace{-4ex}
  \caption{Variation of fitted value of $\ddg$ with
    $N_{\rm min}$ when correction term is fitted.
    The line of best fit to the final six values is shown,
    and our final estimate is plotted on the $y$-axis.}
  \label{fig:DG}
\end{minipage}
\end{center}
\end{figure}

We plot estimates for the amplitudes $\dde$ and $\ddg$ in
Figs~\ref{fig:DE} and \ref{fig:DG}. 
For $\dde$, we see smooth
convergence, and some visible upwards curvature leads us to estimate a
value of $\dde = 1.3112(3)$ which is a little above the linear extrapolation
shown.
For $\ddg$, the curvature over the plot range shown is somewhat larger,
and it is difficult to tell to what extent the trend has really turned
upwards (estimates with successive values of $\Nmin$ are strongly
correlated and so it is necessary to be careful not to read too much
into ``trends'' from just a few data points). Thus in this case our
central estimate of $\ddg = 0.21849(2)$ is somewhat below the linear extrapolation, to guard
against the possibility that the upward trend is due to statistical
noise. N.B., we expect the curvature to decrease as we reach the large
$N$ limit, as unfitted corrections to scaling become relatively smaller
in magnitude, and this is another reason to think that the actual
curvature may be somewhat less than it appears to the eye.

We calculate the ratio of amplitudes, with our confidence interval
calculated by treating
them as statistically independent quantities, obtaining $\dde/\ddg =
6.0012(14)$.
This confirms our expectation that the amplitude ratio assumes the same
value as for simple random walks.

We now seek to calculate the amplitude ratio $\dde/\ddg$ directly from
the ratio $\avresq_N/\avrgsq_N$, where the corresponding data is found
in Table~\ref{tab:rdata}. First we note that
\begin{align}
      \frac{\avresq_N}{\avrgsq_N} &= 
      \frac{D_{\rm E} N [\log (N/a)]^{1/4} \left(1 -
    \frac{17\log(4\log(N/a))+31}{64\log(N/a)} + \cdots \right)}
    {D_{\rm G} N [\log (N/a)]^{1/4} \left(1 -
    \frac{17\log(4\log(N/a))+97/3}{64\log(N/a)} + \cdots \right)}
    \\ &= \frac{D_{\rm E}}{D_{\rm G}} \left(1 + \frac{1}{48\log(N/a)} +
    \cdots \right).
\label{eq:deoverdg}
\end{align}

\begin{figure}[htb]
\begin{center}
\begin{minipage}{0.45\textwidth}
\begin{center}
    \includegraphics[width=1.0\textwidth]{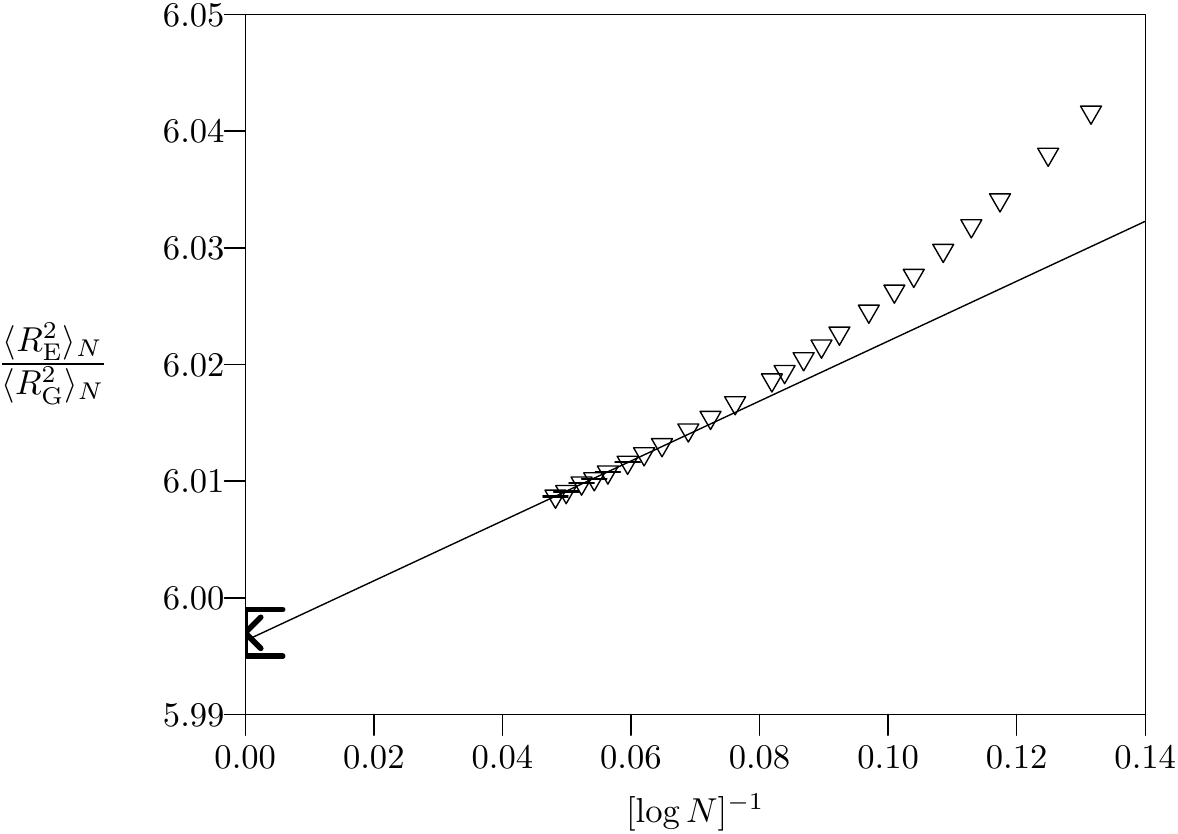}
\end{center}
\vspace{-4ex}
  \caption{Variation of $\avresq_N/\avrgsq_N$ with
    $N$.
    The line of best fit to the final six values is shown,
    and our final estimate is plotted on the $y$-axis. 
    All data up to $N=10^9$ are shown, and
    no error bars are
    visible because they are vanishingly small on the scale of the
    plot.}
  \label{fig:DEoverDG}
\end{minipage}
\hspace{2em}
\begin{minipage}{0.45\textwidth}
\begin{center}
    \includegraphics[width=1.0\textwidth]{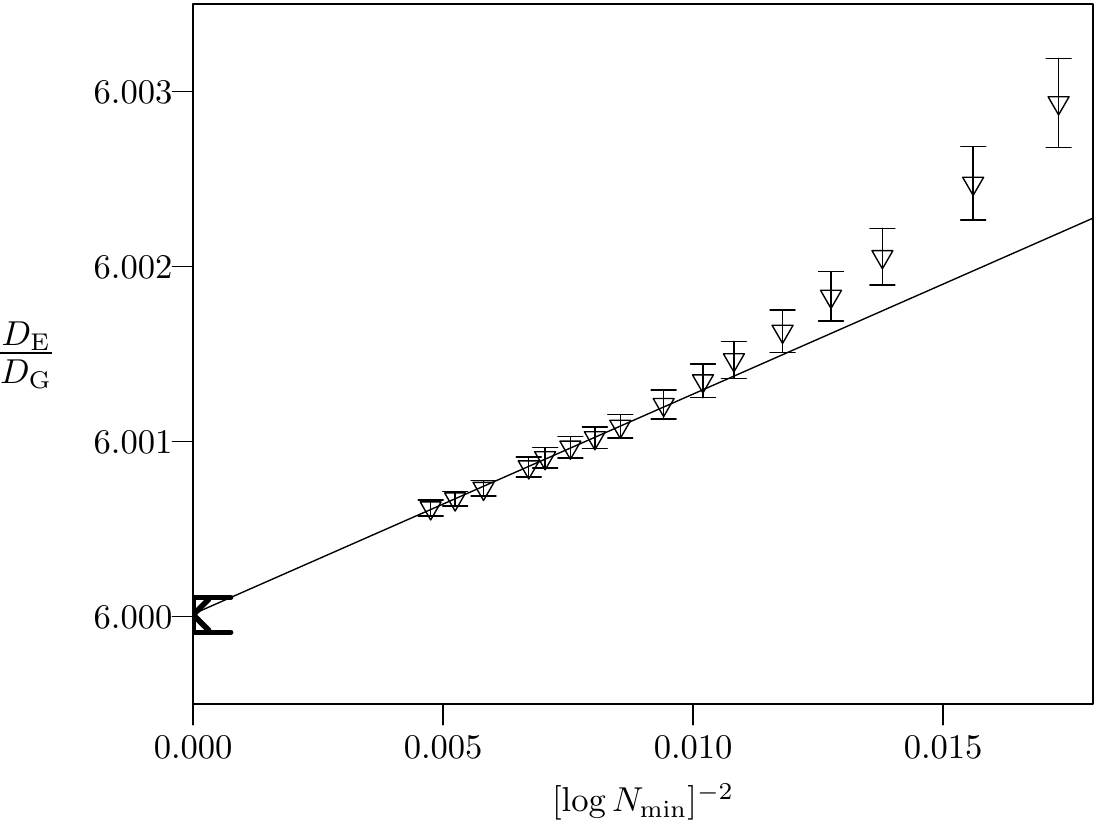}
\end{center}
\vspace{-4ex}
  \caption{Variation of fitted value of $\dde/\ddg$ with
    $N_{\rm min}$ when correction term is fitted.
    The line of best fit to the final six values is shown,
    and our final estimate is plotted on the $y$-axis.}
  \label{fig:DEoverDGcorrection}
\end{minipage}
\end{center}
\end{figure}

In Fig.~\ref{fig:DEoverDG} we plot the ratio
$\avresq_N/\avrgsq_N$ versus $(\log N)^{-1}$; from the equation above we
expect that the ratio will converge linearly to $D_{\rm E}/D_{\rm G}$,
and by performing a linear extrapolation we obtain the estimate  
$D_{\rm E}/D_{\rm G} = 5.997(2)$.

We now fit the correction term in Eq.~\ref{eq:deoverdg},
allowing the parameter $a$ to vary as before. For these fits, the
reduced $\chi^2$ value declines from 70 when $\Nmin = 2000$, to 1.7 when
$\Nmin = 2\e{6}$, indicating that there is still a degree of systematic
error due to neglected corrections to scaling even for the largest
values of $\Nmin$. Extrapolating the fits we obtain the estimate 
$D_{\rm E}/D_{\rm G} = 6.00001(10)$ which is consistent with the
expected value of 6 from the simple random walk, and considerably more
accurate than the value of $6.0012(14)$ which was obtained from the ratios of the
independently estimated amplitudes.

Finally, we consider the behavior of $f$, the probability of a pivot being
successful, as a function of $N$.

In~\cite{Madras1988PivotAlgorithmHighly}, it is argued that a good first
approximation for the probability that a pivot on a walk of $2N$ steps
is successful, is the probability that two independently chosen $N$-step
walks are self-avoiding when they are concatenated. For SAWs in
dimension $d \neq 4$, it is believed that $c_N \sim N^{\gamma-1} \mu^N$,
with $\gamma = 43/32 = 1.34375$ for
$d=2$~\cite{Nienhuis1982ExactCriticalPoint}, $\gamma = 1.15695300(95)$
for $d=3$~\cite{Clisby2017Scale-freeGammaSAWsArxiv}, and $\gamma = 1$
for $d \geq
5$~\cite{Hara1992SelfAvoidingWalk,Hara1992LaceExpansionSelf}. This
argument suggests that $f$ is of $O(N^{1-\gamma})$, but in practice it
is observed that $f = O(N^{-p})$ with $p \approx 0.19$ for $d=2$
(c.f. $\gamma-1 \approx 0.344$) and with $p \approx 0.11$ for $d=3$
(c.f. $\gamma-1 \approx 0.157$). Thus it seems that for $d=2$ and $d=3$
this heuristic is overly pessimistic and the two parts of the walk are
less likely to intersect than independently chosen SAWs which are
concatenated.

For $d \geq 5$ the above argument strongly suggests that the probability of a pivot
move being successful should be $O(1)$, although to the best of our
knowledge this has not been tested.

But, it is less clear what should happen for $d=4$, and hence we will
attempt to extract information about the asymptotic behavior of $f$ from
the data. Firstly, the probability that 
two independently chosen $N$-step
walks are self-avoiding when they are concatenated is exactly
$c_{2N}/c_N^2$. From Eq.~\ref{eq:cn} we can calculate the
asymptotic behavior of this ratio: 
\begin{align}
    \frac{c_{2N}}{c_N^2} &= \frac{A \mu^{2N} [\log (2N/a)]^{1/4} \left(1 -
    \frac{17\log(4\log(2N/a))-3}{64\log(2N/a)} + \cdots \right)}{
        \left(A \mu^N [\log (N/a)]^{1/4} \left(1 -
        \frac{17\log(4\log(N/a))-3}{64\log(N/a)} + \cdots
        \right)\right)^2}
        \\ &= \frac{1}{A}[\log (N/a)]^{-1/4} \left(1 + \cdots \right).
        \label{eq:cnratio}
\end{align}
This expression, together with the heuristic argument given above,
implies that $f$ should be no larger than $O([\log N]^{-1/4})$ for $d=4$.

Now, it is possible to calculate correction terms for
Eq.~\ref{eq:cnratio}, but because the probability of a pivot move being
successful is, at best, only loosely related to $c_{2N}/c_N^2$, there
would not be any point in doing so. Instead, we fit $f$ with a
statistical model which is inspired by the functional form of
Eq.~\ref{eq:cnratio}, given by:
\begin{align}
        f &= [\log (N/a)]^{-\lambda} \left(1 + O([ \log N ]^{-1}) \right).
\end{align}

We perform fits of the data for $f$ in Table~\ref{tab:fdata} for $a$ and
$\lambda$, finding that the reduced $\chi^2$ of the fits declines from
66 when $\Nmin = 2000$, to 2.1 when $\Nmin = 2\e{6}$. Clearly there are
still significant systematic errors even for the largest value of
$\Nmin$, but the fact that the reduced $\chi^2$ is steadily declining,
presumably towards 1, suggests that the statistical model is correct,
and that deviations from the model are due to large unfitted correction
terms. The assumption that the first neglected term was of magnitude
$O([ \log N ]^{-1})$ appears reasonable based on the plot, which appears
to a good approximation to be behaving linearly as the fitted values
approach the $y$-axis. Extrapolating the trend results in the estimate
$\lambda = 0.257(2)$.

\begin{figure}[!htb]
  \begin{center}
    \includegraphics[width=9cm]{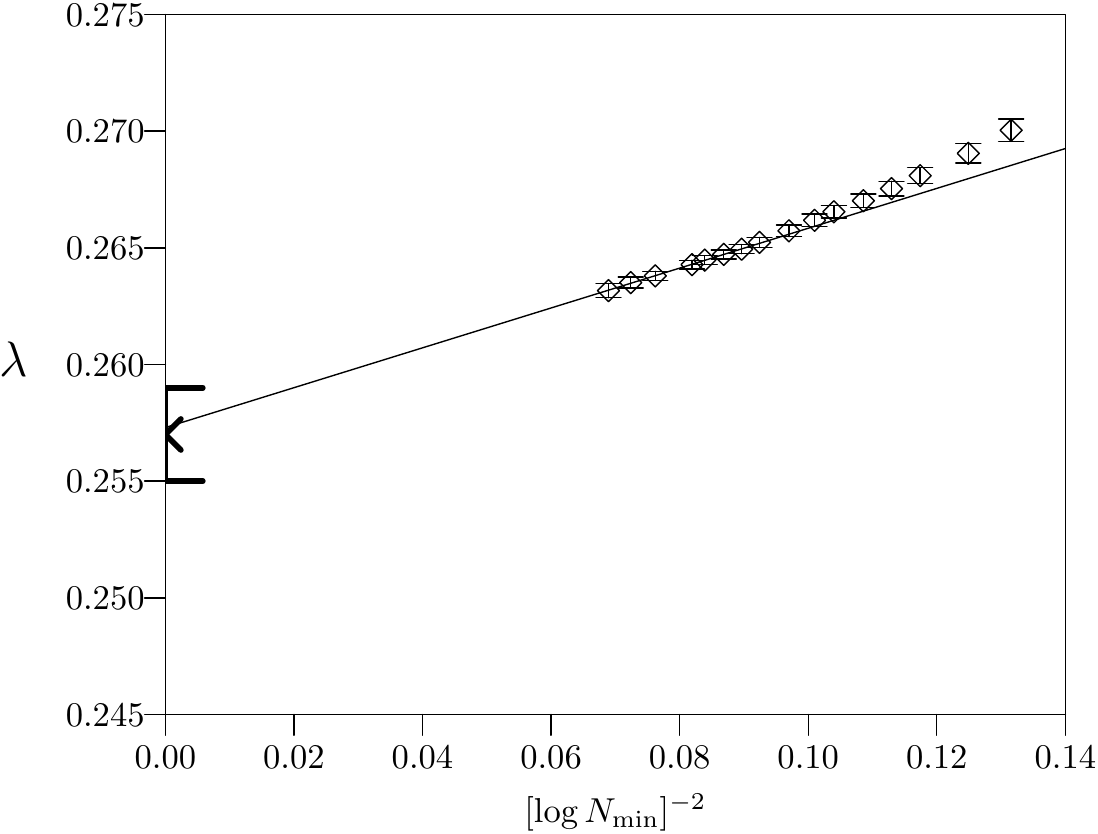}
  \end{center}
    \caption{Variation of $\lambda$ with $N_{\rm min}$ from fitting data for $f$.}
  \label{fig:lambda}
\end{figure}

If the previous heuristic argument holds in a weakened sense, and a pivot move is at least
as likely to be successful as the probability that two independent SAWs
can be concatenated, then this would imply that $\lambda \leq 1/4$.
Since $\lambda$ is very close to $1/4$, and we are not able to imagine
any plausible reason for there to be fine tuning with $\lambda$ 
close to but different from $1/4$, we make the conjecture that
$\lambda = 1/4$ exactly.

As was the case for our leading order fits of $\avresq$ and $\avrgsq$,
our estimated value is somewhat different from the conjectured exact
value, but as there are likely to be strong corrections to
scaling it is perfectly reasonable to suppose that if we had a better
statistical model (as we do for $\avresq$ and $\avrgsq$) then we would
obtain an 
estimate for $\lambda$ that is closer to $1/4$.

\section{Discussion and conclusion}
\label{sec:conclusion}

We found that model fits far poorer in for four-dimensional SAWs than
for comparable computer experiments involving SAWs in three
dimensions~\cite{Clisby2016HydrodynamicRadiusForSAWs,Clisby2017Scale-freeGammaSAWsArxiv}.
In those cases, fitting the leading correction-to-scaling term (or
eliminating it by creating an improved observable) resulted in
``perfect'' fits with reduced $\chi^2$  approximately 1 for $\Nmin$ of the
order of thousands, whereas here we found that although fits were
acceptable there were still significant systematic errors even for
$\Nmin = 2\e{6}$. This is perhaps not surprising given the occurrence of
logarithmic corrections, but it does makes the task of extracting
reliable parameter estimates considerably more difficult.

Nonetheless, by designing the computer experiment to collect high
quality data for $N$ as high as $10^9$, we have found that we are able
to get a good handle on these logarithmic corrections. This was
especially so when correction-to-scaling terms were included in the
fits, but we had modest success in our analyses for the leading exponent
of $\avresq$ and $\avrgsq$, $\kappa$, and the exponent for the
probability of a successful pivot move, $\lambda$, even when only the
leading behavior was fitted.

Our best estimate of $\kappa = 0.2516(14)$ is consistent with the
renormalization group estimate of $\kappa = 1/4$. If further evidence
were needed, the coincidence of our best estimate of the ratio
$\dde/\ddg = 6.00001(10)$ with the simple random walk value of 6 is
confirmation that four-dimensional SAWs are indeed at the boundary
of the simple random walk universality class.

We should mention that the parameter $a$, found in
Eqs~\ref{eq:cn}--\ref{eq:rg2}, which was expected to be a constant,
turned out to vary between different choices of fits, and choices of
observables. This is perhaps not terribly surprising given the
logarithmic corrections to scaling, but does warrant further
investigation.

We studied the behavior of the pivot algorithm for four-dimensional
SAWs. We found that the probability of a pivot move being
successful, $f$, for a SAW of length $N$, is consistent with $f =
O([\log N]^{-1/4})$, and conjectured that this relation is exact.

Finally, one of our key motivations for studying four-dimensional SAWs
was to act as a test-case for attacking problems with logarithmic
corrections to scaling, of which the $\theta$-transition in three
dimensions is another example.  Although the capacity for 
quantitative understanding of
the logarithmic corrections for four-dimensional SAWs would seem to be a
necessary pre-condition for coming to grips with the
$\theta$-transition, it is not sufficient.  This is because the pivot
algorithm for SAWs is remarkably efficient, whereas at the present time
no comparably efficient method exists for sampling interacting
self-avoiding walks (ISAWs) in the vicinity of the $\theta$-transition.
If progress is to be made for the quantitative understanding of the
$\theta$-transition, then significant improvements will need to be made
in the efficiency of Monte Carlo sampling algorithms.

\section*{Acknowledgements}
Support from the Australian Research Council under the Future
Fellowship scheme (project number FT130100972) and Discovery scheme
(project number DP140101110) is gratefully acknowledged.

\appendix

\section{Data}
\label{sec:data}

\renewcommand{\arraystretch}{1.1}
\begin{table}[!ht]
    \caption{Monte Carlo estimates of $\avresq_N$, $\avrgsq_N$, and $\avresq_N / \avrgsq_N$.}
\label{tab:rdata}
\begin{center}
\begin{tabular}{rlll} 
\hline
    \multicolumn{1}{r}{$N$} & \multicolumn{1}{c}{$\avresq_N$} &
    \multicolumn{1}{c}{$\avrgsq_N$}& \multicolumn{1}{c}{$\avresq_N / \avrgsq_N$} \tstrut \bstrut \\
\hline
    2000 & 3.9910993(66)\ee{3} & 6.605999(11)\ee{2} & 6.0416285(53) \\
    3000 & 6.083070(11)\ee{3} & 1.0074613(18)\ee{3} & 6.0380186(57) \\
    5000 & 1.0332096(20)\ee{4} & 1.7122817(33)\ee{3} & 6.0341098(62) \\
    7000 & 1.4636643(31)\ee{4} & 2.4265458(50)\ee{3} & 6.0318840(65) \\
    10000 & 2.1161594(47)\ee{4} & 3.5095129(78)\ee{3} & 6.0297809(70) \\
    15000 & 3.2158126(78)\ee{4} & 5.335114(13)\ee{3} & 6.0276366(73) \\
    20000 & 4.325970(11)\ee{4} & 7.178500(18)\ee{3} & 6.0262867(78) \\
    30000 & 6.567591(18)\ee{4} & 1.0901342(30)\ee{4} & 6.0245715(82) \\
    50000 & 1.1105327(33)\ee{5} & 1.8439156(55)\ee{4} & 6.0226874(89) \\
    70000 & 1.5689845(49)\ee{5} & 2.6056040(83)\ee{4} & 6.0215771(96) \\
    100000 & 2.2624262(76)\ee{5} & 3.757878(13)\ee{4} & 6.020488(10) \\
    150000 & 3.428582(12)\ee{5} & 5.695894(21)\ee{4} & 6.019392(11) \\
    200000 & 4.603711(17)\ee{5} & 7.649043(29)\ee{4} & 6.018676(11) \\
    500000 & 1.1756555(50)\ee{6} & 1.9539824(85)\ee{5} & 6.016715(12) \\
    1000000 & 2.386946(10)\ee{6} & 3.968017(17)\ee{5} & 6.015463(12) \\
    2000000 & 4.842315(23)\ee{6} & 8.051226(39)\ee{5} & 6.014382(14) \\
    5000000 & 1.2322250(65)\ee{7} & 2.049226(11)\ee{6} & 6.013123(15) \\
    10000000 & 2.495866(14)\ee{7} & 4.151237(24)\ee{6} & 6.012343(17) \\
    20000000 & 5.052437(38)\ee{7} & 8.404463(65)\ee{6} & 6.011613(21) \\
    50000000 & 1.282394(10)\ee{8} & 2.133490(18)\ee{7} & 6.010780(24) \\
    100000000 & 2.592982(23)\ee{8} & 4.314303(40)\ee{7} & 6.010201(25) \\
    200000000 & 5.240702(49)\ee{8} & 8.720211(85)\ee{7} & 6.009834(27) \\
    500000000 & 1.327686(20)\ee{9} & 2.209457(35)\ee{8} & 6.009104(43) \\
    1000000000 & 2.680993(46)\ee{9} & 4.461847(80)\ee{8} & 6.008707(49) \\
\hline
\end{tabular}
\end{center}
\end{table}

\begin{table}[!ht]
    \caption{Monte Carlo estimates of the probability of a pivot move
    being successful, $f$, and the mean CPU time per pivot attempt.}
\label{tab:fdata}
\begin{center}
\begin{tabular}{rll} 
\hline
    \multicolumn{1}{r}{$N$} & \multicolumn{1}{c}{$f$} & 
    \multicolumn{1}{c}{CPU time per pivot attempt ($\mu s$)} \tstrut \bstrut \\
\hline
    2000 & 0.65753599(32) & \hspace{2cm} \phantom{1}4.6474(19) \\
    3000 & 0.64883702(35) & \hspace{2cm} \phantom{1}5.1224(20) \\
    5000 & 0.63857892(37) & \hspace{2cm} \phantom{1}5.7122(24) \\
    7000 & 0.63221031(39) & \hspace{2cm} \phantom{1}6.1387(26) \\
    10000 & 0.62576662(42) & \hspace{2cm} \phantom{1}6.6230(30) \\
    15000 & 0.61879528(44) & \hspace{2cm} \phantom{1}7.1649(33) \\
    20000 & 0.61406233(47) & \hspace{2cm} \phantom{1}7.7895(40) \\
    30000 & 0.60766748(50) & \hspace{2cm} \phantom{1}8.3581(42) \\
    50000 & 0.60003595(54) & \hspace{2cm} \phantom{1}9.3899(52) \\
    70000 & 0.59524796(57) & \hspace{2cm} 10.2674(60) \\
    100000 & 0.59036053(60) & \hspace{2cm} 11.1021(59) \\
    150000 & 0.58502890(65) & \hspace{2cm} 12.1095(66) \\
    200000 & 0.58138186(66) & \hspace{2cm} 12.7724(72) \\
    500000 & 0.57043938(75) & \hspace{2cm} 14.9183(88) \\
    1000000 & 0.56277669(76) & \hspace{2cm} 17.2602(92) \\
    2000000 & 0.55558165(82) & \hspace{2cm} 19.317(11) \\
    5000000 & 0.54670734(89) & \hspace{2cm} 22.200(13) \\
    10000000 & 0.54042239(95) & \hspace{2cm} 24.754(15) \\
    20000000 & 0.5344713(12) & \hspace{2cm} 26.809(21) \\
    50000000 & 0.5270577(13) & \hspace{2cm} 30.642(25) \\
    100000000 & 0.5217689(14) & \hspace{2cm} 33.540(27) \\
    200000000 & 0.5167230(14) & \hspace{2cm} 36.676(32) \\
    500000000 & 0.5103988(22) & \hspace{2cm} 40.506(83) \\
    1000000000 & 0.5058602(24) & \hspace{2cm} 50.426(65) \\
\hline
\end{tabular}
\end{center}
\end{table}

\FloatBarrier

\end{document}